\def\spose#1{\hbox to 0pt{#1\hss}}
\def\approxlt{\mathrel{\spose{\lower 3pt\hbox{$\sim$}}
	\raise 2.0pt\hbox{$$<$$}}}
\def\approxgt{\mathrel{\spose{\lower 3pt\hbox{$\sim$}}
	\raise 2.0pt\hbox{$>$}}}

\def\multleft#1{\hbox to size{\vbox {\halign {\lft{##}\cr #1}}\hfill}\par}
\def\multright#1{\hbox to size{\vbox {\halign {\rt{##}\cr #1}}\hfill}\par}

\def\today{\ifcase\month\or January\or February\or March\or April\or May\or
      June\or July\or August\or September\or October\or November\or December\fi
      \space\number\day, \number\year}
\def\$<${\thinspace}
\def\s{\hbox{\phantom{5}}}	

\def\boxit#1{\vbox{\hrule\hbox{\vrule\kern3pt\vbox{\kern3pt
          #1 \kern3pt}\kern3pt\vrule}\hrule}}

\def\cm{{\rm\thinspace cm}}

\def\erg{{\rm\thinspace erg}}
\def\eV{{\rm\thinspace eV}}

\def\K{{\rm\thinspace K}}
\def\keV{{\rm\thinspace keV}}
\def\km{{\rm\thinspace km}}
\def\kpc{{\rm\thinspace kpc}}
\def\Lsun{\hbox{$\rm\thinspace L_{\odot}$}}

\def\Mpc{{\rm\thinspace Mpc}}
\def\Msun{\hbox{$\rm\thinspace M_{\odot}$}}
\def\pc{{\rm\thinspace pc}}

\def\s{{\rm\thinspace s}}
\def\yr{{\rm\thinspace yr}}
\def\sr{{\rm\thinspace sr}}

\def\ergpcmpspsr{\hbox{$\erg\cm^{-2}\s^{-1}\sr^{-1}\,$}}
\def\ergpcmsqps{\hbox{$\erg\cm^{-2}\s^{-1}\,$}}

\def\ergps{\hbox{$\erg\s^{-1}\,$}}

\def\kmps{\hbox{$\km\s^{-1}\,$}}

\def\Msunpsqkpc{\hbox{$\Msun\kpc^{-2}\,$}}

\def\Msunpyr{\hbox{$\Msun\yr^{-1}\,$}}
\def\pcm{\hbox{$\cm^{-3}\,$}}

\def\psqcm{\hbox{$\cm^{-2}\,$}}

\def\kmpspMpc{\hbox{$\kmps\Mpc^{-1}$}}

\def\approxlt{\mathrel{\spose{\lower 3pt\hbox{$\sim$}}
        \raise 2.0pt\hbox{$<$}}}
\def\approxgt{\mathrel{\spose{\lower 3pt\hbox{$\sim$}}
        \raise 2.0pt\hbox{$>$}}}

\documentstyle[psfig]{mn}
\begin{document}
\hsize=6truein

\title{Molecular hydrogen emission in Cygnus A}

\author[]
{\parbox[]{6.in} {R.J.~Wilman$^1$, A.C.~Edge$^2$, R.M.~Johnstone$^1$, C.S.~Crawford$^1$ and A.C.~Fabian$^{1}$\\ \\
\footnotesize
1. Institute of Astronomy, Madingley Road, Cambridge CB3 0HA. \\
2. Department of Physics, University of Durham, Durham, DH1 7LE. \\ }}

\maketitle

\begin{abstract}
We present J, H and K-band spectroscopy of Cygnus A, spanning 1.0--2.4$\mu$m in the rest-frame and hence several rovibrational H$_{2}$, H recombination and [FeII] emission lines. The lines are spatially extended by up to 6\kpc~from the nucleus, but their distinct kinematics indicate that the three groups (H, H$_{2}$ and [FeII]) are not wholly produced in the same gas. The broadest line, [FeII]$\lambda1.644$, exhibits a non-gaussian profile with a broad base ($\rm{FWHM} \simeq 1040$\kmps), perhaps due to interaction with the radio source. Extinctions to the line-emitting regions substantially exceed earlier measurements based on optical H recombination lines.

Hard X-rays from the quasar nucleus are likely to dominate the excitation of the H$_{2}$ emission. The results of Maloney, Hollenbach \& Tielens~(1996) are thus used to infer the total mass of gas in H$_{2}$~v=1--0~S(1)-emitting clouds as a function of radius, for gas densities of $10^{3}$ and $10^{5}$\pcm, and stopping column densities $N_{\rm{H}}=10^{22}-10^{24}$\psqcm. Assuming azimuthal symmetry, at least $2.3 \times 10^{8}$\Msun~of such material is present within 5\kpc~of the nucleus, if the line-emitting clouds see an unobscured quasar spectrum. Alternatively, if the bulk of the X-ray absorption to the nucleus inferred by Ueno et al.~(1994) actually arises in a circumnuclear torus, the implied gas mass rises to $\sim 10^{10}$\Msun. The latter plausibly accounts for $10^{9}$\yr~of mass deposition from the cluster cooling flow, for which $\dot{M} \simeq 10$\Msunpyr within this radius. 
\end{abstract}


\section{INTRODUCTION}
X-ray observations show that the gas within the inner few 100\kpc~of many clusters of galaxies is cooling from temperatures of $10^{7}-10^{8}$\K~at rates of up to several hundred solar masses per year. If persistent for $10^{8}$\yr~or more, the inferred subsonic inflow of cooling material-- the {\em cooling flow} (CF) (see Fabian~1994 for a review)-- would deposit in excess of $10^{10}$\Msun~of cool gas around the central cluster galaxy (CCG). Whilst the presence of such large amounts of cold material has been deduced from soft X-ray absorption (White et al.~1991, Allen et al.~1993, Allen et al.~1995, Allen \& Fabian~1997), its detection in emission has proved problematic and the ultimate fate of the gas remains uncertain. The bulk of it is clearly not turned into visible stars, but the role of dust in obscuring any star formation may be important (Allen~1995; Crawford et al.~1999), and recent sub-millimetre work in fact suggests star formation rates of up to 100\Msunpyr~(Edge et al.~1999). 

There is also a strong link between the cooling time in the central 30\kpc~of a cluster and the presence of optical line-emission and/or a CCG radio source: Peres et al.~(1998) show that for central cooling times below $3 \times 10^{9}$\yr, 90 per cent of the CCGs exhibit line emission, and 95 per cent contain a radio source; for comparison, in CFs with longer central cooling times, the fractions are 5 and 20 per cent, respectively. Despite the uncertainty over the mechanism behind the optical line emission, the surrounding ICM clearly plays an important role. There have recently been attempts to study the emission from cooler phases of the gas, such as molecular hydrogen which emits through thermal excitation at $\sim 2000$\K. Jaffe and Bremer~(1997) detected H$_{2}$~v=1-0~S(1) emission in the inner few \kpc~of three CF CCGs, but not in any radio galaxies outside strong CFs at similar radio powers and redshifts. They found that the emission is too luminous to be due to material simply passing through $\sim 2000$\K~whilst cooling from higher temperatures, and implicated suprathermal electrons deep in the molecular core of the cloud as the reheating mechanism. Falcke et al.~(1998) also observed strong H$_{2}$ emission associated with radio galaxies in CFs, and speculated that it is due to radio-jet interactions with the ICM or to the injection of material from infalling spiral galaxies. Krabbe et al.~(2000) presented H and K-band spectral imaging of NGC 1275, the cD galaxy at the centre of the Perseus cluster, a 200\Msunpyr~CF. They cited the central concentration of the H$_{2}$ emission, its relatively low and uniform temperature (1500-3000\K) and AGN-related excitation mechanism as evidence against the CF hypothesis. But such properties are in fact entirely consistent with the expectation that much of the gas is too cold ($<100$\K) to emit significant H$_{2}$ emission, and is only observed when excited in some way, in this case by the AGN itself.

\begin{table*}
\caption{Log of observations}
\begin{tabular}{|llllll|} 
UT Date &  $\lambda-\lambda$  &  Resolution $\dagger$  &   Seeing$\ddagger$ &   Integration  &   Slit \\ 
(1999)	&   ($\mu$m)	      &  (\kmps~at FWHM)       &  (arcsec)      &    time (min)  & PA (deg) \\ 
        &                     &                     &                &                       &       \\ 
Sep 2	& 1.875-2.490 (K)	      &     570             &   1.0	     &    64   &      0 \\
Sep 3   & 1.345-1.961 (H)        &     880             &   1.2          &    48   &      0 \\
Oct 7	& 1.128-1.435 (J)	      &     610             &   1.5 $\star$  &    52   &      0  \\ 
\end{tabular}

$\dagger$ At centre of the wavelength range. \\
$\ddagger$ FWHM of standard star spatial profile along the slit. \\
$\star$ The night was non-photometric. \\

\end{table*}

We have recently begun a near-infrared study of the H$_{2}$ emission in a sample of 28 CF CCGs, chosen to be the most optically line luminous in the complete spectral study of 217 ROSAT-selected clusters of Crawford et al.~(1999). In this paper we present J, H and K band spectra of Cygnus A, which was observed early in the programme. At redshift $z=0.056$, it is by some considerable margin the most powerful radio source in the local universe and, like the radio galaxy 3C295, similar in power to the $z \geq 1$ FRII radio sources. It resides in a moderately rich, cooling flow cluster with a mass deposition rate of $\sim 250$\Msunpyr~(Reynolds \& Fabian~1996); X-ray observations also reveal an absorbed power-law component from the quasar nucleus (Ueno et al.~1994), which further manifests itself through scattered broad emission lines (Antonucci, Hurt \& Kinney~1994; Ogle et al.~1997).

This paper is structured as follows: the observations and data reduction are described in section~2, followed in section~3 by a discussion of the extended emission in the lines of molecular hydrogen, the hydrogen recombination lines of Pa$\alpha$, Pa$\gamma$, Br$\gamma$ and Br$\delta$, and [FeII]$\lambda\lambda1.258,1.644$; in section 4, we make deductions about the gas content and molecular excitation mechanisms within the central few \kpc~of Cygnus A, in the context of the current understanding of the CF and the obscured quasar. The cosmological parameters $H_{0}=50$\kmpspMpc~and $q_{0}=0.5$ are adopted throughout, yielding a spatial scale of 1.5\kpc~arcsec$^{-1}$ at the redshift of Cygnus A.

\section{OBSERVATIONS AND DATA REDUCTION}
The observations were taken with the CGS4 spectrograph on the United Kingdom Infrared Telescope (UKIRT) in September and October 1999, as shown in the 
observation log in Table 1. The $256\times256$ InSb array, 40~l/mm grating and
300~mm focal length camera were in place, yielding a spatial scale of 0.61 
arcsec per pixel and spectral resolutions of 610, 880 and 570\kmps~FWHM (in 
the J, H and K bands, respectively) with a 2-pixel wide slit aligned north-south. The NDSTARE mode was used along with the conventional object-sky-sky-object
nodding pattern, thus obviating the need for separate bias and dark current 
frames and permitting the computation of an external error for each pixel; it 
also reduces the required accuracy of the flat-field frame. Atmospheric 
absorption features were removed by ratioing with some of the main sequence F 
stars tabulated at the telescope which were calibrated against photometric 
standards (except for the J-band exposure acquired under non-photometric 
conditions, and for which an approximate flux calibration was obtained by 
matching the continuum level with that in the H-band). On- and off-line data 
reduction was performed using version V1.3-0 of the Portable CGS4 Data 
Reduction package available through Starlink. Row-by-row spectra were 
extracted from the fully-reduced spectral images and converted to ASCII 
format for use with the emission-line fitting package QDP/PLT (Tennant 1991). 
Emission lines were fitted with gaussian components atop a linear continuum.

\section{RESULTS}
\begin{figure}
\psfig{figure=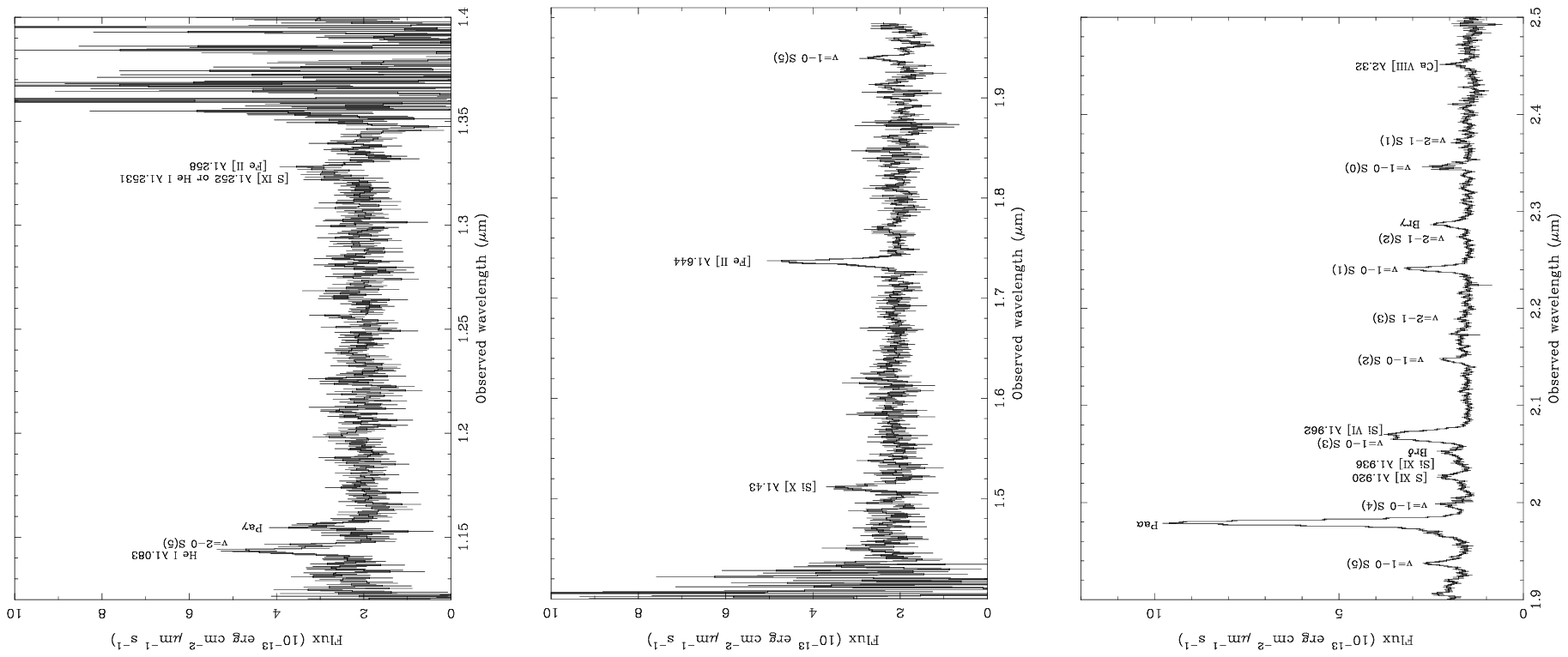,width=0.48\textwidth,angle=270}
\caption{\normalsize J, H and K-band spectra of the nucleus of Cygnus A, plotted with 1$\sigma$ error bars obtained from the data reduction; Pa$\beta$ lies in the atmospheric absorption band at 1.354$\mu$m.}
\label{fig:cygAspec}
\end{figure}

Spectra from the nuclear row of Cygnus A are shown in Fig.~\ref{fig:cygAspec}.
Many of the lines were previously identified by Ward et al.~(1991) and 
Thornton et al.~(1999) in earlier UKIRT spectra. The blue shoulder on 
Pa$\alpha$ is plausibly due to HeII$\lambda1.8639$ (using case B theory and 
the HeII$\lambda4686$ flux in Osterbrock \& Miller~1975, its predicted flux is 
$3\times10^{-16}$\ergpcmsqps) and also perhaps to HeI$\lambda\lambda1.8691,1.8702$ (although we cannot predict their fluxes); Thornton et al.~(1999), however, ascribed it entirely to very high excitation H$_{2}$ O-series lines. Using the list of coronal lines in Ferguson et al.~(1997), we identify several new such lines, viz. [Ca VIII]$\lambda 2.32$, [Si X]$\lambda 1.43$ and [S IX]$\lambda 1.252$, in addition to those previously known between 2.0 and 2.1$\mu$m. As an alternative to [S IX]$\lambda 1.252$, the blue component of the doublet near 1.325$\mu$m could instead be HeI$\lambda1.2531$ and/or H$_{2}$~v=2-0~Q(4), but if it were the latter, the absence of the Q(1), Q(3) and Q(5) lines of this series (which should be much stronger under thermal excitation) would be hard to explain. In the present data, many of the identified emission lines extend significantly beyond the instrumental point spread function (PSF), as described below.

\subsection{Spatially extended line fluxes}
\begin{figure}
\psfig{figure=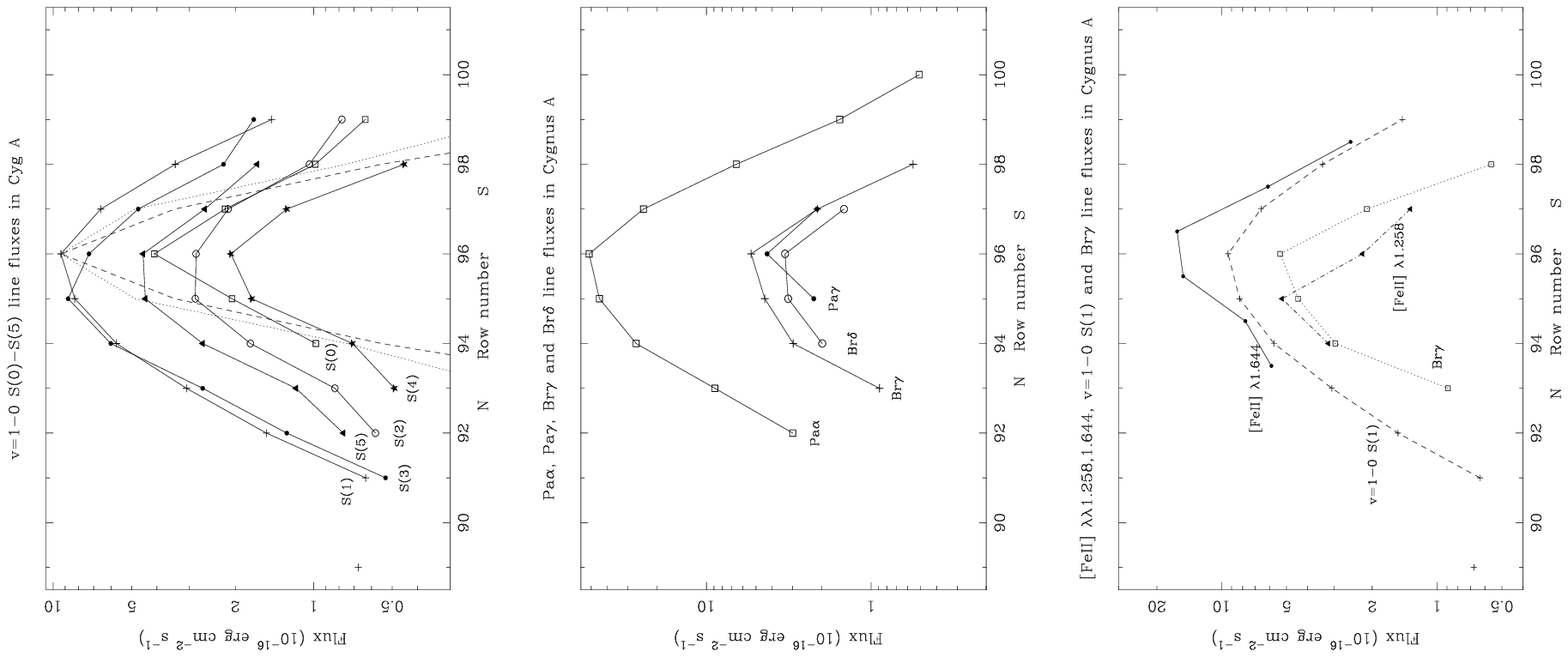,width=0.48\textwidth,angle=270}
\caption{\normalsize Emission line fluxes as a function of row number; the peak in the underlying continuum in row 96 of the K-band exposure is taken to mark the position of the nucleus (the H-band points, i.e. [FeII]$\lambda1.644$, have thus been shifted by 0.5 pixel, as explained in the text; e.g. the line at row 96 on the chip is shown here at 96.5). The unlabelled curves in the upper plot are PSFs in the intervals 1.9--1.95$\mu m$ (dotted line) and 2.15--2.25$\mu m$ (dashed line). The spatial scale is 0.6 arcsec per pixel.}
\label{fig:extlines}
\end{figure}

Fig.~\ref{fig:extlines} shows the emission line fluxes as functions of position along the slit. The nuclear light of the H-band exposure falls on two rows of the chip, so the [FeII]$\lambda 1.644$ points in Figs.~\ref{fig:extlines} and \ref{fig:kinem} have been shifted by $+0.5$ pixel. All the lines are spatially extended, especially to the north, and by up to 6.3\kpc~in the case of v=1-0~S(1). There are notable variations in the line ratios, e.g. in S(1)/S(3), and in ratios involving [FeII]~$\lambda\lambda1.258,1.644$, Br$\gamma$ and v=1-0 S(1), which we discuss in section~4. We caution, however, that there are systematic uncertainties in the S(3) flux in rows 96--98 owing to the blending of this line with [Si VI]$\lambda$1.962.

\subsection{Kinematics}
\begin{figure}
\psfig{figure=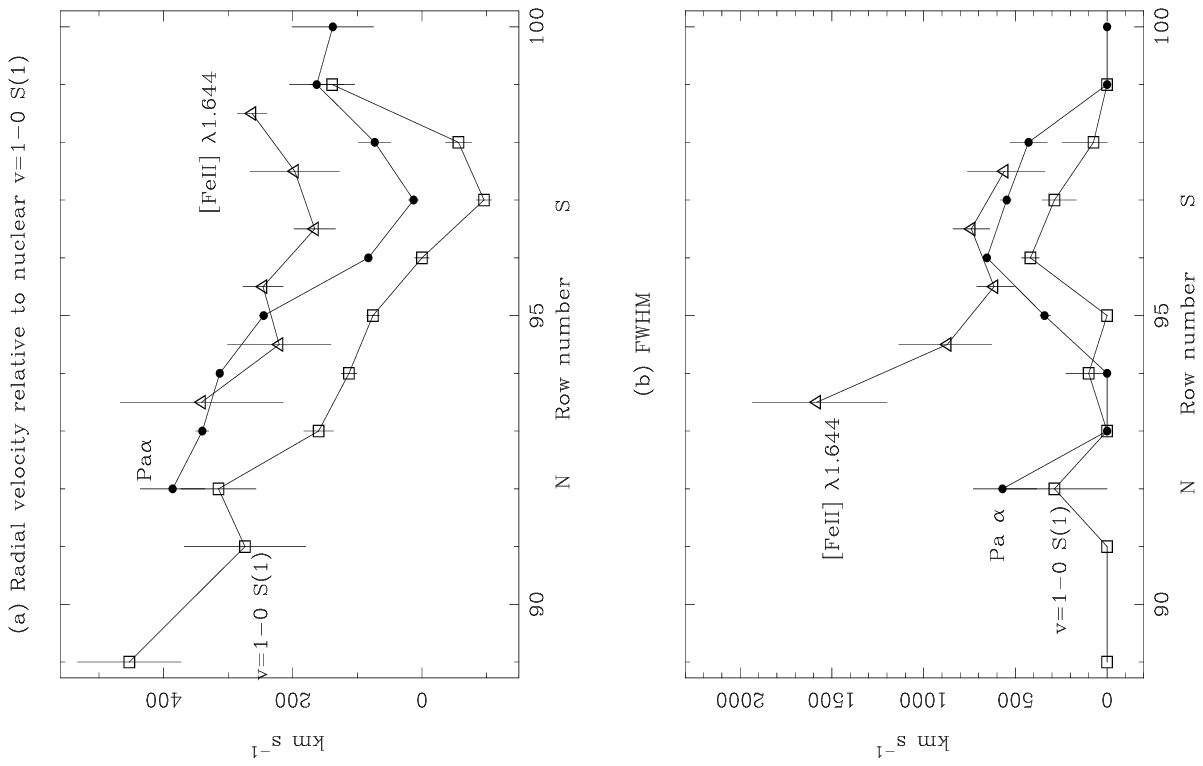,width=0.48\textwidth,angle=270}
\caption{\normalsize (a) The radial velocities of the H$_{2}$~v=1-0~S(1), Pa$\alpha$ and [FeII]$\lambda1.644$ as functions of row number, with respect to the v=1-0~S(1) emission in the nucleus (row 96). (b) The FWHM of these lines, after correction for the instrumental resolution (points at FWHM=0 \kmps~and without error bars represent unresolved lines). The spatial scale is 0.6 arcsec per pixel.}
\label{fig:kinem}
\end{figure}

Fig.~\ref{fig:kinem} shows the variations along the slit in the radial velocity and FWHM of H$_{2}$~v=1-0~S(1), Pa$\alpha$ and [FeII]$\lambda$1.644. That the kinematics of the three lines differ from one another shows that they are not entirely produced in the same gas. The radial velocity curves of H$_{2}$~v=1-0~S(1) and Pa$\alpha$, showing a rise to the north to $\simeq 400$\kmps~and an initial decrease to the south before increasing again, are at least qualitatively consistent with the behaviour of [OIII]$\lambda5007$ and [NII]$\lambda6584$ found by Stockton et al.~(1994) at the same PA. Using a narrow slit offset from the nucleus by various amounts, they found that these lines split into two components separated by 300\kmps~1~arcsec west of the nucleus and that the gradient of the velocity curve also changes sign there, due to an emission component located north-west of the nucleus. Also from optical lines, Tadhunter et al.~(1994) found relatively large line widths (200--600\kmps~FWHM) along the radio axis (PA 105 degrees), but only small radial velocity shifts ($<200$\kmps), although Tadhunter~(1991) had earlier found a high velocity cloud, moving at 1500--1800\kmps~from the galaxy rest frame and associated with the north-west emission line component. 

The S(1) line has FWHM$\simeq420$\kmps~on nucleus (in agreement with the measurement by Thornton et al.~1999), and is significantly broader to the south than to the north. In rows 96--98 where S(3) is strongly blended with [Si VI]$\lambda$1.962, the velocity width of the former was set to that of the S(1) line in a two gaussian fit; the fitted width of [Si VI] was found to be $\simeq 930$\kmps~FWHM in rows 96--98 (consistent with Ward et al.'s value of $1200\pm300$\kmps), $\simeq 550$\kmps in row 95 and unresolved further to the north. Thornton et al.'s low value of 582\kmps~for the FWHM of [SiVI] on nucleus and high value of 552\kmps~for the S(3) line (much higher than the other v=1-0~S lines) demonstrates that leaving both line widths free in fitting such a blend can lead to errors (their value of 1.2 for the v=1-0~S(3)/S(1) line ratio is thus too high). We also note that on nucleus (row 96), the para-hydrogen lines v=1-0~S(0), S(2) and S(4) are all significantly narrower than the ortho-hydrogen lines, v=1-0~S(1), S(3) and S(5): S(1), S(3) and S(5) have a mean width of 410\kmps~FWHM, whereas S(0) and S(2) are unresolved and the S(4) width $\simeq 200$\kmps~FWHM; this implies that in this row the ortho and para lines may in part be produced in distinct regions with different (and at least in part, non-thermal) ortho/para population ratios.

\begin{figure}
\psfig{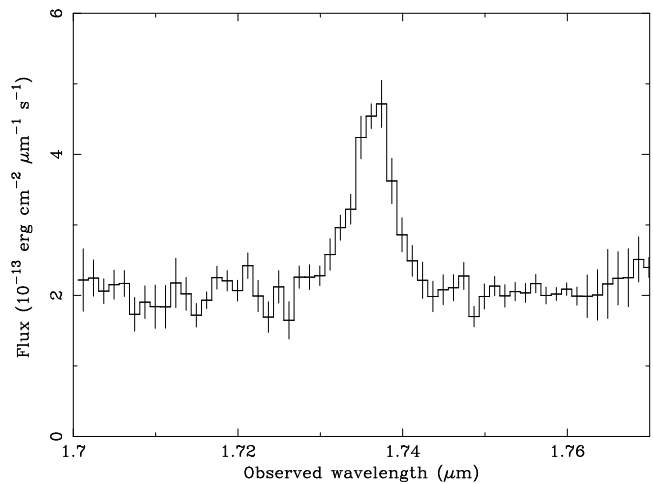}
\caption{\normalsize The distinctly non-gaussian profile of the [FeII]$\lambda1.644$line in row 95 of the H-band exposure; in a two component fit, the broader component (FWHM$\simeq1040$\kmps) contains 70~per cent of the flux, the rest being in the unresolved core.}
\label{fig:fe2row95}
\end{figure}

The fitted width of Pa$\alpha$ is $\simeq 630$\kmps~FWHM on nucleus, and generally higher to the south than to the north (its blue shoulder was modelled with an additional component). The [FeII]$\lambda1.644$ line is as broad as Pa$\alpha$ on nucleus but shows an off-nucleus increase, which may be due to an additional narrow component on nucleus. This is shown in Fig.~\ref{fig:fe2row95} for row 95, where the broad (FWHM=1040\kmps) base contains 70 per cent of the flux, with the narrow (unresolved) core providing the rest. 

We speculate that the broader, spatially-extended [FeII]$\lambda1.644$ component results from shocks (e.g.~from the radio source), which plausibly produce the larger line-width and raise the gas phase Fe abundance by evapouring it from dust grains. In this regard, we note that Forbes \& Ward~(1993) attributed the [FeII] emission in starburst and Seyfert galaxies to shock-excitation in supernova remnants and radio jets, respectively. Like the latter authors, Simpson et al.~(1996) also found that the [FeII] emission in active galaxies is more tightly correlated with radio power than is the H recombination emission, but concluded that photoionization is the dominant [FeII] production mechanism, although $\sim 20$~per cent of it may be shock-excited. Krabbe et al.~(2000) also observed a markedly non-gaussian [FeII]$\lambda1.644$ profile in NGC 1275, leading them to suggest that it is produced in two distinct regions. We also note that Whittle~(1992) found that the FWHM of the [OIII]$\lambda5007$ line in Seyfert galaxies without linear radio sources is largely determined by gravity (correlating well with the amplitude of the galaxy rotation curve); Seyferts with luminous linear radio sources, however, are observed to have significantly broader lines, due to acceleration by the radio source.

\begin{figure}
\psfig{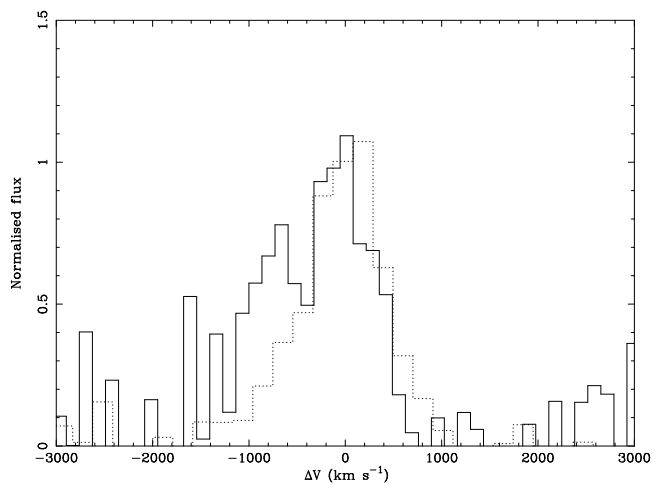}
\caption{\normalsize Profiles of the [FeII]$\lambda1.258$ (solid) and [FeII]$\lambda1.644$ (dotted) lines in velocity space, after subtraction of the fitted linear continuum and scaled to the peak flux of the fitted line. The velocity centroid of each line is taken from a gaussian fit. It is difficult to infer whether the broad base to [FeII]$\lambda1.644$ is also present on [FeII]$\lambda1.258$, due to the presence of either [S IX]$\lambda 1.252$ and/or HeI$\lambda 1.2531$ on the blue wing of [FeII]$\lambda1.258$.}
\label{fig:fe2vel}
\end{figure}

In the J-band, the FWHM of Pa$\gamma$ is $430^{+340}_{-430}$\kmps~(with 1$\sigma$ errors), although with the errors it is consistent with the value of 630\kmps~FWHM observed for Pa$\alpha$. The fitting of the [FeII]$\lambda1.258$ line is complicated by the line on its blue wing which, as mentioned earlier, is either HeI$\lambda 1.2531$ or [SIX]$\lambda1.252$: on nucleus, the fitted FWHM is $430^{+230}_{-430}$\kmps~if the widths of the two lines are left free, or $570^{+210}_{-150}$\kmps~if they are constrained to be the same. These ranges are consistent with the FWHM of [FeII]$\lambda1.644$ on nucleus, but it is not possible from these data to infer whether or not [FeII]$\lambda1.258$ also has a broad base (since the two lines originate from the same upper level, they should have identical profiles in the absence of any differential extinction -- recall from Table 1, however, that the instrumental resolution is slightly poorer in the H-band). This is illustrated by the line profile comparison in Fig.~\ref{fig:fe2vel}; the lines have been assumed to have the same velocity centroid, although the fitted position of [FeII]$\lambda1.258$ appears to be blueshifted by $\sim 200$\kmps~from [FeII]$\lambda1.644$ (but the HeI$\lambda 1.2531$ and/or [SIX]$\lambda1.252$ would tend to reduce the fitted centroid of [FeII]$\lambda1.258$, and the relative wavelength calibration between 1.25 and 1.64$\mu m$ is only accurate to $\sim 100$\kmps). It should also be noted that the slit could have been in a slightly different position during the J-band observation, which was taken under non-photometric conditions with poorer seeing.

\subsection{Reddening}
On nucleus, we measure Br$\gamma$/Pa$\alpha=0.107\pm0.007$ which translates into an extinction of A$_{\rm{V}}\simeq9.6$~mag, assuming an intrinsic line ratio of 0.083 (case B) and the near-infrared interstellar extinction curve of Landini et al.~(1984). For comparison, Ward et al.~(1991) deduced A$_{\rm{V}}\simeq 1.2\pm0.15$~mag using Pa$\alpha$/H$\beta$, and Tadhunter et al.~(1994) found A$_{\rm{V}}\simeq 3$ from H$\alpha$/H$\beta$. There are two possible reasons for the discrepancies: firstly, Br$\gamma$ may be contaminated by some HeI lines on its blue wing (although their expected strengths are not known); alternatively, the obscuring material may not be in the form of a uniform foreground screen (e.g. it may have variable thickness, or be mixed with the gas), in which case decrements between different optical and near-infrared lines would not yield the same A$_{\rm{V}}$. For completeness, we note that Br$\gamma$/Pa$\alpha=0.112\pm0.013$ in row 94, equivalent to A$_{\rm{V}}\simeq 11.5$~mag; in row 98, Br$\gamma$/Pa$\alpha=0.0685\pm0.033$, consistent with the case B value and therefore with no extinction.

Drawing upon the J-band data, we measure Pa$\gamma$/Pa$\alpha=0.083\pm0.019$ on nucleus, which translates into A$_{\rm{V}}=5.7$~mag; in rows 95 and 97 we infer A$_{\rm{V}}=8.2$ and 5.4~mag, respectively. We may also compute the extinction to the [FeII]-emitting region. Intrinsically, we expect a fixed [FeII]$\lambda1.258/\lambda1.644$ ratio of 1.35 (Nussbaumer \& Storey~1988), but find on nucleus that [FeII]$\lambda1.258/\lambda1.644 \simeq 0.3$, implying A$_{\rm{V}}=12$~mag; the extinction is at least as large in row 97, but somewhat less in row 95 (A$_{\rm{V}} \simeq 9$~mag). As noted at the end of section~3.2, however, the presence of the broad component to [FeII]$\lambda1.644$ which may be lacking from [FeII]$\lambda1.258$, suggests that this component is subject to {\em more} extinction than the values just calculated, and the {\em narrow} component which dominates [FeII]$\lambda1.258$ to less (i.e. a single foreground extinction screen is not a satisfactory model). We also reiterate the caveat concerning the possibility that the slit may have been at a slightly different position during the J-band observation.

Concerning the extinction to the dust continuum-emitting regions, Imanishi \& Ueno~(2000) measured the optical depth of the 9.7$\mu$m silicate absorption feature and found $\tau_{9.7} \sim 1$ (equivalent to $A_{\rm{V}} \sim 9-19$~mag if the dust properties resemble those of Galactic dust). That this figure is much less than the $A_{\rm{V}} \sim 150$~mag to the $3\mu$m-emitting region implies, they argue, that there is a dust temperature gradient, and their models suggest that this requires a dusty torus with an inner region of $<10$\pc. It is of interest that their $A_{\rm{V}} \sim 9-19$~mag to the 10$\mu$m continuum emission site does not significantly exceed our inferred extinctions. For convenience, we summarise the latter in Table~2.

\begin{table}
\caption{Extinctions to line-emitting regions}
\begin{tabular}{|lll|} 
Line ratio & reference$\dagger$ &  A$_{\rm{V}}$         \\ 
           &           &  on nucleus (mag) $\ddagger$       \\ \\
Br$\gamma$/Pa$\alpha$ &                        & 9.6 \\
Pa$\gamma$/Pa$\alpha$ &                        & 5.7 \\
Pa$\alpha$/H$\beta$   & Ward et al.~(1991)     & 1.2 \\
H$\alpha$/H$\beta$    & Tadhunter et al.~(1994)  & 3   \\
FeII$\lambda 1.258 / \lambda 1.644$ &    & 12  \\ \\ \\
\end{tabular}
$\dagger$ This work unless stated. \\
$\ddagger$ Calculated using a foreground screen extinction model.  \\
\end{table}

\section{INTERPRETATION}

\subsection{LTE Comparison}
From the findings of section~3 we wish to infer the properties of the line-emitting H$_{2}$ and the associated excitation mechanism. If the v=1-0~S(1) emission is thermally excited at a vibrational temperature of 2000\K, we may follow Scoville et al.~(1982) and deduce the mass of warm H$_{2}$ at this temperature; using the total v=1-0~S(1) luminosity of $(5.7\pm0.3)\times 10^{40}$\ergps~(integrated along the slit), we find $M$(warm~H$_{2})=(2.4\pm0.1)\times10^{5}$\Msun, much less than the total mass which the cooling flow would have deposited, but the bulk of that material would in any case be much too cold ($<100$\K) to produce thermally-excited S(1) emission. Following Jaffe \& Bremer~(1997), we may also compute the mass cooling rate, $\dot{M}$, by multiplying the v=1-0~S(1) luminosity in photon~s$^{-1}$ by the mass of the H$_{2}$ molecule and dividing by the fraction of the total cooling rate at 2000\K~in this line, usually estimated at between 2 and 10 per cent: we find $\dot{M}>3 \times 10^{4}$\Msunpyr, over 100 times greater than the X-ray mass deposition rate. We are thus driven to the same conclusions as Jaffe \& Bremer, namely that the observed line emission is not simply due to material passing through $\sim 2000$\K~whilst cooling from 10$^{7}$\K~and that an additional heat source is therefore required. 

\begin{figure}
\psfig{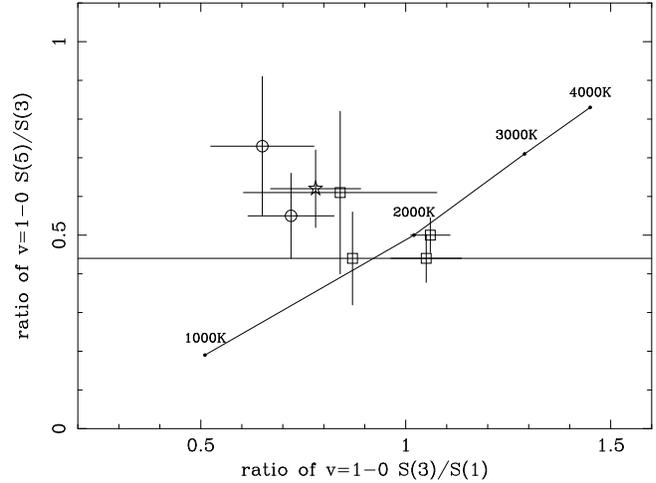}
\caption{\normalsize Comparison of the ratios of several H$_{2}$ v=1-0 S-series lines with the predictions for thermal excitation at various temperatures; the star denotes the nucleus (row 96), circles rows to the south (rows 97 and 98), and squares rows to the north (rows 92-95).}
\label{fig:lte}
\end{figure}

Clues to the nature of the latter may be gleaned from Fig.~\ref{fig:lte}, which shows for each row flux ratios involving the v=1-0~S(1), S(3) and S(5) lines; the upper levels of these transitions lie 6956, 8365 and 10341\K~above the ground state, respectively, giving sensitivity to a broad range of temperatures. It is apparent from this figure that much of the emission is {\em not} thermally excited in high density gas at a single temperature, especially in the nucleus and to the south of it. The points deviate from the LTE prediction in the sense expected for emission from gas at a range of temperatures, where the emission in the lower excitation lines would have a relatively greater contribution from lower temperature gas. We repeat the caution that these ratios involve fluxes in the S(3) line which in rows 96, 97 and 98 is blended with [Si VI], and it is these points which lie farthest from the LTE curve. The ratios of some para-hydrogen lines, viz. v=1-0~S(4)/S(2) and S(2)/S(0), also deviate from the LTE prediction, thereby attesting to the effect being real. These deviations indicate that the lines are either produced in low density gas (below the relevant critical densities of $\sim 10^{5}$\pcm) or that they are excited by a non-thermal mechanism, or perhaps both. Non-thermal excitation by suprathermal electrons liberated by hard X-ray photoionization within the cold, highly molecular core of the cloud was implicated by Jaffe \& Bremer~(1997) on the basis of the high v=1-0~S(1)/H$\alpha$ ratio, and we consider it in detail below. 

\subsection{Comparison with an X-ray dissociation region model}
Cold clouds deposited by cooling flows and immersed in the radiation field of the surrounding ICM, can produce molecular line emission in two distinct regions. Part of the emission can be generated via thermal excitation immediately beyond the ionized skin of the cloud, peaking where the molecular fraction first reaches $\sim 10^{-3}$ and the temperature $\sim 2000$\K. At greater depths, the temperature drops rapidly towards that of the microwave background and the gas becomes highly molecular (Ferland, Fabian \& Johnstone~1994). At these depths, primary and secondary electrons generated in situ by hard X-ray photons can excite appreciable H$_{2}$ emission (this contrasts with the situation in a highly-ionized gas where primary photoelectrons rapidly thermalize through collisions with other electrons). Only by including this latter contribution to the v=1-0~S(1) emission were Jaffe \& Bremer able to reconcile the high observed v=1-0~S(1)/H$\alpha$ ratio with the calculations of Ferland, Fabian \& Johnstone. The most detailed calculations of the emission from X-ray-irradiated molecular gas are those by Maloney, Hollenbach \& Tielens~(1996) (hereafter MHT) and we proceed to apply them to Cygnus A.
 
As discussed by MHT, the dominant parameter controlling the physical conditions is $H_{\rm{X}}/n$, the ratio of the X-ray energy deposition rate per particle, $H_{\rm{X}}$, to the gas density, $n$. The former is the integral over energy of $F(E)\sigma_{\rm{pa}}$, where $F(E)$ is the local photon energy flux per unit energy interval, and $\sigma_{\rm{pa}}$ the photoelectric cross-section per hydrogen atom. For an X-ray source with a power-law spectrum of photon index $\Gamma=2$ and 1-100\keV~luminosity $L_{\rm{X}}=10^{44}L_{\rm{44}}$\ergps, located 100r$_{2}$\pc~from the gas cloud, MHT give the following expression for $H_{\rm{X}}$:

\begin{equation}
\hspace{1cm} H_{\rm{X}} \sim 7 \times 10^{-22} L_{\rm{44}}~ r_{2}^{-2}~ N_{22}^{-1}~~\ergps
\end{equation}
where $N_{\rm{att}}=10^{22}N_{22}$\psqcm~is the hydrogen column density attentuating the X-ray flux; they assume a column of at least $N_{\rm{att}}=10^{21}$\psqcm~in order to exclude the normal UV photon-dominated region at the cloud surface. Figs.~6a and 6b of MHT show the surface brightness, $\Im(H_{x}/n)$, (from a cloud of column density $10^{22}$\psqcm) in the [FeII]$\lambda 1.644$, Br$\gamma$, v=1-0~S(1) and v=2-1~S(1) lines as functions of $H_{\rm{X}}/n$ at the cloud centre, for densities $n=10^{3}$ and $10^{5}$\pcm. These figures can be used to compute the total luminosity, $L_{i}$, in line $i$ from an area $A_{\rm{cl}}$ of cloud-face exposed to the continuum source, emitted in the column density interval $(N_{1,22}-N_{2,22}) 10^{22}$\psqcm:

\begin{equation}
\hspace{1cm} L_{i}=A_{cl} \int_{N_{1,22}}^{N_{2,22}} \Im_{i}(H_{X}/n) \ dN_{22}\end{equation}
using eqn.~(1) to evaluate $H_{\rm{X}}/n$ as a function of $N_{22}$. Thus eqn.~(2) can be used in conjunction with the observed v=1-0~S(1) luminosity in each row to derive $A_{cl}$ as a function of $N_{2,22}$ and cloud density, and thence the {\em total} mass of gas in the cloud, not merely that in warm H$_{2}$.

The radio sources in the CCGs considered by Jaffe \& Bremer~(1997) were weak enough to be negligible in comparison with the CF itself as an X-ray excitation source. This is not the case for Cygnus A, which houses a powerful obscured quasar: the {\em Ginga} spectrum of Ueno et al.~(1994) exhibits an absorbed power-law component, with a photon index $\Gamma\simeq2$, a line of sight absorption $N_{\rm{H}} \simeq 3.7 \times 10^{23}$\psqcm~and an unabsorbed 2-10\keV~luminosity of $1.1\times 10^{45}$\ergps. The only circumstances under which this component would not be the dominant source of X-rays at the radii of interest would be if the line of sight from the quasar nucleus to the extended gas were obscured by the above column density or more, whilst the X-rays from the cooling flow were unattenuated. The off-nuclear increases in the v=1-0~S(1)/[FeII]$\lambda1.644$ and v=1-0~S(1)/Br$\gamma$ ratios (Fig.~\ref{fig:extlines}) are, with reference to Figs.~6a and 6b of MHT, consistent with the expected off-nuclear decrease in $H_{\rm{X}}/n$~from a nuclear excitation source. In fact, since only the {\em narrow} component of [FeII]$\lambda1.644$ is produced in the same region as v=1-0~S(1), the v=1-0~S(1)/[FeII]$\lambda1.644$ ratio relevant in this context increases more sharply than Fig.~\ref{fig:extlines} indicates.

\subsubsection{The mass of off-nucleus v=1-0~S(1)-emitting clouds}

\begin{figure}
\psfig{figure=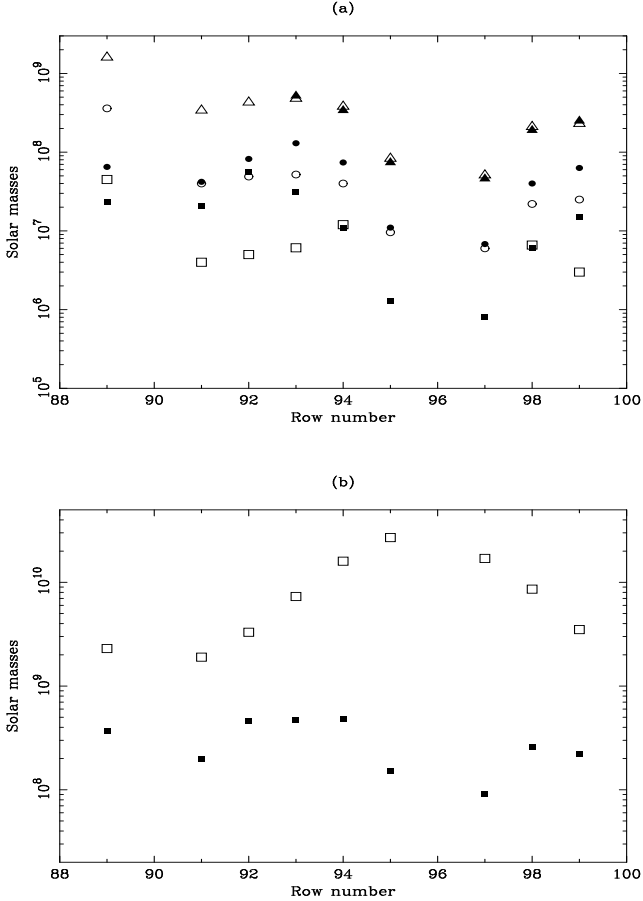,width=0.48\textwidth,angle=270}
\caption{\normalsize The {\em total} mass of gas in the v=1--0 S(1)-emitting clouds, as functions of off-nucleus row number (the nucleus is in row 96), for densities of $n=10^{3}$\pcm~(open symbols) and $10^{5}$\pcm~(filled symbols), under the following two scenarios: (a) when the emitting clouds are exposed to an unattenuated quasar spectrum, with the square, circular and triangular symbols denoting integration of eqn.~(2) to stopping column densities of $N_{\rm{H}}=10^{22}, 10^{23}$ and $10^{24}$\psqcm, respectively; (b) when the emitting clouds see a quasar spectrum absorbed by $N_{\rm{H}} = 3.7 \times 10^{23}$\psqcm~due to the nuclear torus; see text for details. Each pixel is 0.6 arcsec across.}
\label{fig:gmass}
\end{figure}

Following the procedure outlined above, we compute the {\em total} mass of gas in the v=1-0~S(1)-emitting clouds (ionized+neutral+molecular) in each off-nucleus pixel. We assume that all of the emission is from gas at the projected radius of the pixel, which strictly speaking means that the calculated masses are lower limits on the true values projected on to each pixel. We perform the calculation for two limiting cases: firstly, where the gas is exposed to the {\em unobscured} quasar X-ray spectrum, obtained by extrapolation to $\sim 100$\keV~of the power-law observed by Ueno et al.~(1994); secondly, for the case where the X-ray spectrum is absorbed by $3.7 \times 10^{23}$\psqcm~{\em on nucleus}, in a circumnuclear torus. For each of the two densities modelled by MHT, we integrate eqn.~(2) from $N_{1,22}=0.1$ to $N_{2,22}=$1,~10 and 100; in some cases the minimum value of $H_{\rm{X}}/n$ plotted by MHT ($10^{-29.4}$\ergps~cm$^{3}$) is reached before $N_{2,22}=100$, in which case a point is not shown for this column. 
The results for the case where the gas sees an unobscured spectrum, shown Fig.~\ref{fig:gmass}a, demonstrate that the inferred mass is relatively insensitive to the assumed gas density. There is a stronger dependence on the stopping column density, especially close to the nucleus, where the bulk of the emission is produced near the thermal-excitation peak within $N_{2,22}=1$. Reducing the stopping column below $N_{2,22}=1$ (but still $N_{2,22}>0.1$), however, does not substantially reduce the implied gas masses, so the lowest points in each pixel of Fig.~\ref{fig:gmass} may be considered approximate lower limits. Noting that each pixel of the detector covers a projected area of 1.7\kpc$^{2}$ in Cygnus A, we use Fig.~\ref{fig:gmass}a to place lower limits on the projected gas mass as a function of radius: in rows 95 and 97 (i.e. within 1\kpc~of the nucleus) $\Sigma \geq 6 \times 10^{5}$\Msunpsqkpc~(this does {\em not} include the spatially-unresolved nuclear emission), and in rows 91--94, 98 and 99 (i.e. within roughly 5\kpc~of the nucleus) $\Sigma \geq 3 \times 10^{6}$\Msunpsqkpc. If the distribution of this material is azimuthally symmetric, this implies that there are {\em at least} $2.3 \times 10^{8}$\Msun~of H$_{2}$~v=1--0~S(1)-emitting gas clouds within 5\kpc~of the nucleus. We note that the higher mass points in Fig.~\ref{fig:gmass} imply that much more gas could be accommodated without overproducing the flux in this line; and if the points in row 89 are azimuthally-representative, there could be a comparable mass at $r \geq 5$\kpc. Correction of the H$_{2}$~v=1--0~S(1) fluxes for intrinsic extinction would increase the implied masses yet further. 

The fact that the surface mass density is not centrally concentrated in Fig.~\ref{fig:gmass}a indicates that this gas bears no relation to the quasar nucleus, apart from the fact that its emission is excited by it. Rather, we favour an interpretation in which this cold material is part of that deposited by the 250\Msunpyr~CF. In their deprojection analysis of the {\em ROSAT} PSPC data, Reynolds \& Fabian~(1996) found a mass deposition rate of $\simeq 35$\Msunpyr~within the smallest spatial bin (radius 20\kpc); on the assumption that the $\dot{M}(<r) \propto r$ observed at larger radii continues to hold, $\dot{M} \sim 10$\Msunpyr~within 5\kpc. Thus, on a timescale of $10^{9}$ years, the CF would have deposited $\sim 10^{10}$\Msun~within this radius. From Fig.~\ref{fig:gmass}, we see that we can account for this quantity of material without violating constraints on the v=1-0~S(1) line emission only by integrating to a column density of $10^{24}$\psqcm. 

The results for the case where the X-ray spectrum is absorbed on nucleus are shown in Fig.~\ref{fig:gmass}b. In this case, the clouds are exposed to a much weaker X-ray flux, and the emission is produced entirely in the non-thermal regime of Figs.~6a and 6b of MHT. We consider the case where the clouds are much less than $3.7 \times 10^{23}$\psqcm~thick, which results in an X-ray energy deposition rate which is relatively independent of depth within the cloud. The implied gas masses are thus independent of the actual cloud thickness. For the $n=10^{5}$\pcm~case, we calculate a total mass of $1.6 \times 10^{10}$\Msun~within a projected radius of 5\kpc, which accounts for more than $10^{9}$\yr~of mass deposition from the 10\Msunpyr~CF within this radius. For the low density case, $n=10^{3}$\pcm, the much higher mass would clearly violate kinematic constraints.

\subsubsection{The nuclear emission}
In addition, the nuclear v=1-0~S(1) flux can be used to place constraints on the putative circumnuclear torus, which may be responsible for the X-ray absorption towards the nucleus. According to Maloney~(1996), this structure is likely to be $\sim 100$\pc~in size: the detection of broad HI absorption (FWHM$\sim 270$\kmps) towards the nuclear radio source (Conway \& Blanco~1995) argues against a $\sim$\kpc-scale torus; limits on the free-free absorption at 1.4~GHz and the effects of the non-thermal continuum on the spin temperature of the 21\cm~line rule out a pc-scale torus. We noted in section~3.3, however, that Imanishi \& Ueno~(2000) argued in favour of a $<10$\pc-scale torus.

Using a radius of 100\pc~and integrating eqn.~(2) to the X-ray-inferred column density of $N_{2,22}=37$, we find a v=1-0~S(1) surface brightness of 2.3\ergpcmsqps~for $n=10^{5}$\pcm~(for $n=10^{3}$\pcm, the value of $H_{\rm{X}}/n$ in the cloud exceeds the range of the MHT models). In conjunction with the nuclear v=1-0~S(1) luminosity of $2.5\times 10^{40}$\ergps~(corrected for light scattered into neighbouring rows), we deduce a total gas mass of $3.3 \times 10^{6}$\Msun. As seen from the central source, such material would cover 0.9 per cent of the sky. If the density of the torus is closer to $n=10^{3}$\pcm, the molecular fraction will be much lower, leading to a correspondingly larger gas mass and covering factor. As discussed by Maloney~(1996), the failure of Barvainis \& Antonucci~(1994) to detect CO absorption is most simply explained if the torus is wholly atomic. 

Alternatively, if the inner radius of the torus is $<10$\pc, as suggested recently by Imanishi et al.~(2000), $H_{\rm{X}}/n$ would be much higher and outside the range plotted by MHT. For densities of $n=10^{3}$ and $10^{5}$\pcm~the surface brightness of the S(1) line would thus be much lower, leading to a higher gas mass and covering factor. Other excitation mechanisms may, however, be important, as discussed in section 4.3.

Black~(1998) has suggested that if the torus were mostly molecular, the H$^{+}_{3}$ ion may be sufficiently abundant to produce observable ro-vibrational lines in the K-band. He assumed the torus to lie at 300\pc~from the nucleus, and calculated the line intensities for two limiting cases, corresponding to kinetic temperatures of 100 and 1000\K. None of his predicted lines (which would have intensities in the range 0.2--1.0 times that of Br$\gamma$ in Fig.~\ref{fig:cygAspec}) are seen in our spectrum, which also suggests that the torus is not highly molecular.

\subsection{Alternative excitation mechanisms}
We now consider two other excitation mechanisms for the H$_{2}$ emission, namely UV radiation from young stars or the active nucleus itself, and shocks. In low density gas ($n \ll 10^{5}$\pcm), rovibrational H$_{2}$ emission can be produced by direct photoexcitation of excited vibrational states, characterised by strong emission in the v=2-1~S(1) line relative to the v=1-0 lines of lower excitation; in gas above the critical density ($\sim 10^{5}$\pcm), a sufficiently strong UV radiation source can heat the gas to a kinetic temperature of $\sim 1000$\K, leading to collisional excitation of the H$_{2}$ lines (see the detailed calculations of Sternberg \& Dalgarno~1989, hereafter SB89).

We first consider the energetic requirements for UV excitation. The average surface brightness of the v=1-0~S(1) line over rows 89--99 is $2.6 \times 10^{-5}$\ergpcmpspsr~in the Cygnus A frame (values for individual pixels are within a factor of 4 of this value). From the models of SB89, the emission in this line represents no more than 2 per cent of the total 1--4$\mu m$ H$_{2}$ line flux, so the overall H$_{2}$ line surface brightness is at least $1.3 \times 10^{-3}$\ergpcmpspsr; from Figs.~9 and 10 of SB89, this requires a density $n \approxgt 10^{4.7}$\pcm, and a 5--13.6\eV~UV background which is a factor of $\chi \approxgt 10^{3}$ higher than the standard galactic value of $2.5 \times 10^{-4}$\ergpcmpspsr~(Draine~1978). The implied 5--13.6\eV~luminosity is $1.4\times 10^{11}$\Lsun. 

Star-formation is often observed in the central galaxies of CF clusters (Allen~1995; Crawford et al.~1999), so we consider it as a possible source of exciting UV radiation. From {\em HST} WFPC2 imaging, Jackson et al.~(1998) identified several blue condensations within the the spiral structure which is present within a 2 arcsec radius of the nucleus of Cygnus A, and deduced from their colours that they are plausibly due to star-formation within the last Gyr. Their implied luminosity of $3 \times 10^{7}$\Lsun~is, however, insufficient to power the observed H$_{2}$ emission. Furthermore, our north-south slit lies along the nuclear dust-lane and does not encompass any of these condensations (the finite seeing will of course scatter some of their light into the slit). On energetic grounds we cannot completely rule out the possibility that the H$_{2}$ emission is powered by heavily obscured star-formation along the dust-lane, since the monochromatic {\em IRAS} 60$\mu m$ luminosity of Cygnus A, although very weak for its radio lobe power (Barthel \& Arnaud~1996), is $5.25 \times 10^{11}$\Lsun. 

The active nucleus is also a potential source of exciting UV photons. Imaging polarimetry by Tadhunter, Scarrott \& Rolph~(1990) and Ogle et al.~(1997) suggests that there is an extended region which scatters light from a hidden nucleus; its morphology resembles that of the ionization cone seen in the {\em HST} image of Jackson et al.~(1996), with a half opening angle of 55 degrees, oriented approximately along the radio axis. Shaw \& Tadhunter~(1994) performed long-slit spectroscopy to determine the spectral energy distribution (SED) of the featureless blue continuum (FBC) in the nuclear regions, after removing the old galaxy component. In the region covered by our slit, they found a monochromatic surface brightness of $1.4 \times 10^{-5}$\ergpcmpspsr\AA$^{-1}$~at a rest-frame wavelength of 3620\AA, after correction for intrinsic extinction of E(B-V)=0.9~mag; the SED of the FBC steepens below 4200\AA, in a manner which could be consistent with the scattering of quasar light by dust. If $F_{\rm{\lambda}}$ continues to be at least as steep as $\lambda^{-3}$ below 3620\AA, then its 5--13.6\eV~surface brightness will exceed 0.35\ergpcmpspsr, and thus be sufficient to power the observed H$_{2}$ emission. Furthermore, since the FBC is scattered nuclear light, any H$_{2}$ within the ionization cone will be exposed to a stronger UV radiation field (which could, however, be sufficient to destroy the H$_{2}$ altogether).

Thus, on energetic grounds, scattered nuclear UV light or obscured star formation could potentially power the observed H$_{2}$ emission. Both possibilities are, however, largely ruled out by the observed line ratios. As mentioned earlier, the v=2-1~S(1)/v=1-0~S(1) flux ratio is an important diagnostic for UV excitation. The v=2-1~S(1) line is detected only in rows 94 and 96, with v=2-1~S(1)/v=1-0~S(1) flux ratios of $0.23\pm0.06$ and $0.13\pm0.03$, respectively, for which the equivalent collisional (LTE) excitation temperatures, $T_{\rm{12}}$, are 3130 and 2380\K. For UV pumping in low density gas , $T_{\rm{12}} \sim 10^{4}$\K. With the higher density gas ($n \approxgt 10^{4.7}$\pcm) and intense UV radiation field ($\chi \approxgt 10^{3}$) required on energetic grounds, $T_{\rm{12}} \sim 1000$\K~and the v=1-0 transitions would be collisionally-excited at the kinetic temperature of the gas (also $\sim 1000$\K), contrary to the results displayed in Fig.~\ref{fig:lte}. We thus conclude that UV radiation does {\em not} play an important role in the excitation of the H$_{2}$ emission.

Finally we consider shocks as an excitation mechanism, following the method of Draine \& Woods~(1990). The fraction of the kinetic energy dissipated in a shock which is radiated in the v=1-0~S(1) line is $\epsilon(v_{\rm{s}})$, where $v_{\rm{s}}$ is the shock speed. For magnetised molecular clouds, $\epsilon \simeq 0.02$ for $30 \leq v_{\rm{s}} \leq 50$\kmps~(Draine, Roberge \& Dalgarno~1983), decreasing at lower shock speeds because the gas is not hot enough to emit in this line, and at higher shock speeds because the H$_{2}$ is dissociated. For the v=1-0~S(1) slit luminosity of $5.7 \times 10^{40}$\ergps, we thus require a power of $\sim 3\times 10^{42}$\ergps~to be dissipated in shocks. The kinetic energy of $\sim 10^{9}$\Msun~of gas (equal to that which the CF would have deposited within the slit over 1~Gyr), moving with random velocities of $\sim 50$\kmps, is $\sim 3 \times 10^{55}$\erg; it could thus support this rate of dissipation for only $\sim 3 \times 10^{5}$\yr. Since the mass of {\em warm} molecular hydrogen is only $2.4 \times 10^{5}$\Msun~(see section 4.1), only $\sim 10^{-4}$ of the cold material from the CF would be emitting at any one time. As stated in Jaffe \& Bremer~(1997), assuming that it takes 1\yr~for shocked material to return to quiescence, the material must be shocked every $10^{3}-10^{4}$\yr. Based on the shortness of these timescales and on the mismatch between the required random cloud velocities for efficient H$_{2}$ emission ($\sim 50$\kmps)~and the observed line widths, we do not consider shocks to be important in this context. Hard X-ray heating remains the most viable excitation mechanism.

\section{SUMMARY AND CONCLUSIONS}
We have presented J, H and K-band UKIRT CGS4 spectroscopy of Cygnus A, as 
part of a programme to study the H$_{2}$ emission in a sample of cooling flow 
central cluster galaxies. Spanning a rest-frame wavelength range of 1.0--2.4$\mu$m, 
the spectra include the S(0)--S(5) lines of the H$_{2}$~v=1--0 series, 
Pa$\alpha,\gamma$, Br$\gamma,\delta$ and [FeII]$\lambda\lambda$1.258,1.644. The lines are spatially extended by up to 6\kpc~from nucleus. Kinematic 
differences between them indicate that the three types of line emission (H$_{2}$, H recombination and [FeII]) are not produced entirely in the same gas clouds. In particular, much of the [FeII]$\lambda1.644$ line originates in a broad 
component ($\rm{FWHM} \simeq 1040$\kmps), perhaps due to interaction with the radio 
source, but an unresolved core also contributes on nucleus; the H 
recombination lines and those of H$_{2}$ are succesively narrower. Foreground screen extinction models imply A$_{\rm{V}}$=9.6 (from Br$\gamma$/Pa$\alpha$), 5.7 (from Pa$\alpha$/Pa$\gamma$) and 12 mag (from the [FeII] lines), substantially greater than the extinctions deduced by previous workers using optical lines; comparable extinctions are also deduced in some off-nucleus rows.

The relative intensities of various H$_{2}$~v=1-0~S lines deviate from the expectations for thermal (collisional) excitation in gas at a single kinetic temperature. Hard X-rays from the obscured quasar nucleus are found to be the most plausible excitation source for the H$_{2}$ emission; we thus applied the results of MHT on molecular line emission from X-ray dissociation regions to deduce the projected mass density ($\Sigma$) in v=1-0~S(1)--emitting clouds, for gas densities of $10^{3}$ and $10^{5}$\pcm~and stopping column densities between $10^{22}$ and $10^{24}$\psqcm. If the observed variation of $\Sigma$ along the slit is azimuthally representative, it implies a mass of at least $2.3 \times 10^{8}$\Msun~within 5\kpc~of the nucleus. The near uniform variation of $\Sigma$ with radius strongly suggests that this material represents part of that deposited by the cooling flow (for which $\dot{M} \simeq 10$\Msunpyr~inside this radius) rendered visible in this manner by the `searchlight' of the powerful quasar. Moreover, if much of the X-ray absorption to the point source seen by Ueno et al.~(1994) actually arises in a $<100$\pc-scale circumnuclear torus, the implied mass in cold clouds is much larger, $\sim 10^{10}$\Msun, plausibly accounting for $\sim 10^{9}$\yr~of mass deposition from the cooling flow. 

In a subsequent paper we shall apply these techniques to the larger sample of CF CCGs mentioned in section~1.

\section*{ACKNOWLEDGMENTS}
UKIRT is operated by the Joint Astronomy Centre on behalf of the United Kingdom
Particle Physics and Astronomy Research Council. RJW and RMJ acknowledge support from PPARC, and ACE, CSC and ACF thank the Royal Society for support.

{}

\end{document}